# Liquid refractive index sensing independent of opacity using an optofluidic sensor based on diffraction

Zhida Xu[1], Kevin Han[1], Ibrahim Khan[1], Xinhao Wang[1] and G. Logan Liu[1,*]

[1]*Micro and Nanotechnology Laboratory, Department of Electrical and Computer Engineering, University of Illinois at Urbana-Champaign, Urbana, IL 61801, USA*
*Corresponding author: loganliu@illinois.edu*

**Abstract**

Previously, we have implemented a multi-functional optofluidic sensor which can monitor the change of the refractive index and pressure of biofluid simultaneously and can detect free-solution molecular interaction in-situ. Compared to our previous work, this paper has two major improvements proved by both simulation and experiments. One improvement is the broader dynamic measurement range of refractive index by making the diffraction grating with high-index polymer. The other improvement is the separation of refractive index sensing and opacity sensing by using the relative power ratio of diffraction orders for refractive index sensing while using the absolute power for opacity sensing. This simple, compact and low-cost multi-functional optofluidic sensor can be used for in-situ biofluid monitoring, for example, blood monitoring in hemodialysis.

**Introduction**

Monitoring of biofluid plays a critical role in many biomedical applications such as diagnosis of disease. Abundant information can be extracted by probing the physical properties of biofluid including refractive index, density, opacity, conductivity and viscosity. Among those physical properties, refractive index (RI) of biofluid is of particular interest because it contains lot of information like the blood glucose level [1] and label-free molecular interaction [2]. In recent decades, many optical sensors for fluid and biofluid sensing have been developed due to their high accuracy, wide dynamic range, electrical passiveness, repeatability and non-intrusiveness [3] and many of them are specifically for RI measurement. The RI sensing schemes include laser interferometry [4], capillary [5], photonic crystals [6], surface plasmon resonance [7], optical fibers [8] and diffraction gratings [9-11]. Most of those approaches are for RI sensing only and require sophisticated facilities including lens, beam splitter, spectrometers and so on. Even though there exist a few examples that can measure several parameters including refractive index, absorption and temperature simultaneously [12-14], they all require complicated setup and use different sensing schemes for different parameters. Monitoring multiple parameters of biofluid simultaneously is of great significance in the era of big data. For example, in hemodialysis, multiple blood properties including the arterial and venous pressure and the concentration of dialysate and waste have to be constantly monitored [15, 16]. Previously, we have demonstrated a multi-functional optofluidic sensor with an elastomeric 2D grating and a hemispherical fluid chamber [17]. It can use the transmission diffraction pattern for RI sensing while use the reflection diffraction pattern for pressure sensing. The free-solution binding of bovine serum albumin (BSA) and anti-BSA Immunoglobulin G can be detected with this optofluidic sensor. Neither complicated equipment nor strenuous alignment is needed. In this work, we made two major advancements on this optofluidic sensor and the method of interpreting the data. On one hand, we found by numerical simulation that by changing the RI of the grating polymer, we can shift the most sensitive RI sensing range to the RI range of our interest. Then we confirmed the simulation results experimentally by using higher RI polymer for the diffraction grating. On the other hand, while previously we used the diffraction power and angle to monitor the RI change [17], in this work we chose the relative power ratio between different transmission diffraction orders for RI sensing while chose the absolute power for opacity sensing. By interpreting the data in this way, we not only improved the reliability of RI sensing, but also separated the opacity sensing from the RI sensing. Our ultimate goal is to ease and bring down the cost of the clinical monitoring of whole blood in hemodialysis with this multi-functional optofluidic sensor.

**Methods**

The sensing principle of the diffraction optofluidic sensor is schematically demonstrated in Fig. 1(a) and a photograph of the prototype sensing system is shown as Fig. 1(b). The setup merely requires a laser pointer and a screen in addition to the sensing chamber and little alignment is needed. The major sensing element is the 2D diffraction grating in the center of hemispherical fluid chamber. A scanning electron microscope (SEM) image of the 2D pyramids grating is shown in Fig. 1(c). Both the grating and the fluid chamber are made of transparent and colorless polymer. The grating is made by nano-replication process described in details previously by us [18], with a commercially available inverted-pyramids array silicon substrate as the molding template [19]. After the laser beam passes through the diffraction grating, it is diffracted to different directions and can form a diffraction pattern on the screen (Fig. 1(d)). The direct transmission is referred as order 00; the 4 orders closest to the order 00 to its upside, downside, left and right are referred as order 10; the 4 orders on the diagonals outside order 10 are referred to order 11. The transmission diffraction angle is determined by the grating equation $nd\sin(\theta_m) = m\lambda$. For a fixed wavelength λ, grating constant d, and diffraction order m, the diffraction angle $\theta_m$ depends on the RI of the liquid in the hemispherical fluid chamber. The purpose of designing the fluid chamber to be hemispherical is to make sure that all diffracted beams undergo the same distance in the liquid so they get attenuated equally and to make sure that the diffracted laser beams are not deflected at the liquid-polymer interface due to refraction. In our previous work, we have proved that even though we can use the diffraction angle to monitor the change of refractive index n, using diffraction power or grating efficiency to monitor the change of n is a lot more sensitive [17]. The diffracted power was measured with Thorlabs GmbH S130C kit with photodetector and powermeter. We have also demonstrated using the reflection diffraction pattern to monitor the pressure change with the deflectable grating made of the elastomeric polymer polydimethylsiloxane (PDMS) [17].

**Shifting the most sensitive region of RI sensing**

We have shown in [17] that the diffraction pattern will become invisible and the sensitivity will drop if n of liquid goes beyond 1.38 while the grating is made of PDMS(n = 1.4). It is because of the weakening of diffraction efficiency when RI of liquid approaches that of the diffraction grating. As a result, we want to make the diffraction grating with higher-RI polymer for stronger diffraction. To investigate how the sensitivity of RI sensing will be affected by changing the RI of grating, we use finite-difference time-domain (FDTD) method to simulate for the diffraction of 2D pyramids grating with different RI of grating, carried out by the commercial software Lumerical FDTD Solutions. Fig. 2(a) shows the pyramids array in the FDTD model. The wave is incident vertically from the bottom of the pyramids array and the forward diffraction pattern is calculated in the farfield. To make the diffraction pattern symmetrical, we insert two wave sources with same propagation direction and phase but orthogonal polarizations. Fig. 2(b) shows a representative simulated diffraction pattern in the farfield when the ambient RI is 1. Fig. 2 (c-f) show the simulation results of the power of different diffraction orders at different RI of grating and liquid. The simulated power intensity for a certain diffraction power is taken as the locally maximal intensity of electric field for this diffraction order. The intensity is a unit-less number relative to the electric field intensity of incident wave, which is 1 in numeral. Fig. 2(c-e) show the simulated power intensity of order 11, order 10 and order 00 respectively at different RI of liquid and polymer grating. Each curve represents how the power changes with the increase of RI of liquid at one certain RI of grating. One clear trend is that as the RI of grating increases, the curve shifts to the high RI direction. For a fixed RI of grating, the relation between the diffraction power and the RI of liquid is not linear or monotonic. Take grating RI = 1.4 for example, for order 11 (Fig.2(c)) and order 10 (Fig. 2(d)) sharing the same trend, the region most suitable for RI sensing, which is the most linear and highest-sloped region, is from RI = 1.2 to 1.35. As the RI of grating increases to 1.55, this operating RI region shifts to from 1.3 to 1.5. Since most biofluids such as blood serum have the RI range from 1.3 to 1.4 [20], it is reasonable to shift the operating RI region to this range. To find the operating RI region for any grating RI, we find the liquid RI where the slope is the highest for each grating RI and plotted them in Fig. 2(f). Then we can safely say that the most sensitive liquid RI sensing region or operating regions increases as the grating RI increases, but not in a linear relation.

To verify the shift of operating region by experiment, we fabricated the pyramids array grating with PDMS (RI = 1.4) and ultraviolet(UV)-curable polymer(Norland NOA61, RI = 1.56) and prepared sucrose solution with different concentrations as liquid with different RI. The relationship between the concentration and RI of sucrose solution can be found in Cell Biology Laboratory Manual[21]. The dye called brilliant green was mixed with the sucrose solutions for different absorptions to test how the results of refractive index sensing will be affected by the absorption or opacity. The results were plotted in Fig. 3. For order 10 and 11, the powers measured at four diffraction spots were averaged and the standard deviations were calculated. Fig. 3(a-c) show diffracted power of order 11, order 10 and order 00 respectively at the grating RI = 1.4. Fig. 3(d-f) show diffracted power of order 11, order 10 and order 00 respectively at the grating RI = 1.56. Comparing Fig. 3(a,b) with Fig. 3(d,e), we can see for grating RI = 1.4, the power of order 11 and 10 first decrease than increase as liquid RI increases and reach the minimum around liquid RI = 1.4 while for grating RI = 1.56, the power of order 11 and 10 decrease monotonically and almost linearly as liquid RI increases. The measurement results match very well with the simulations data shown in Fig. 2(c,d). However, the sensitivity at grating RI = 1.56 seems to be lower than at grating RI = 1.4 in spite of its better linearity and monotonicity. Let us take the clear solution for order 10(black dots in Fig. 3(a) and (d)) for example. For liquid RI from 1.33 to 1.4, for grating RI = 1.4, the power drops from 21.3 µW to 3.85 µW by 81.92%; for grating RI = 1.56, the power drops from 137.35 µW to 92.525 µW by 32.64%. The relative sensitivity for grating RI = 1.4 is calculated as 1228.2%/RIU while 489.36%/RIU in the liquid RI region from 1.33 to 1.4. As a result, we conclude that by increasing the grating RI from 1.4 to 1.56, we increase the dynamic range but with the tradeoff of low sensitivity.

 Separation of the refractive index and opacity sensing

On Fig. 3 we can see that as the opacity (concentration of brilliant green) of the liquid increases, the diffraction power drops. That means the RI sensing will be interfered by the opacity of the liquid if the absolute diffraction power is used for RI sensing. However, we observed that for the same opacity, order 11, order 10 and order 00 are attenuated almost proportionally equal. So we assume that if the power ratio of 2 different orders is used for RI sensing, the interference by the opacity can be avoided. Actually, that is why we designed the fluid chamber to be hemispherical as shown by Fig. 1(a). The laser beam is diffracted at the center of the hemispherical chamber so different orders travel the same distance before they hit the wall of the hemispherical chamber. As a result, different diffraction orders are attenuated equally in the hemispherical chamber. Considering that order 11 and order 10 share the same trend, we divided the diffraction power of order 11 and order 10 by the power of order 00 and plotted the results in Fig.4. The power ratios of order 11 to order 00 and order 10 to order 00 at different opacity when grating RI = 1.4 were shown in Fig. 4(a) and Fig. 4(b). The power ratios of order 11 to order 00 and order 10 to order 00 at different opacity when grating RI = 1.56 were shown in Fig. 4(c) and Fig. 4(d). As predicted, for either grating RI = 1.4 or 1.56, the curves of power ratio at different opacity are very close to each other, except at the highest opacity(30 µM brilliant green). The reason for the inconsistency at 30 µM brilliant green is possibly attributed to the very dim light intensity at such high opacity, which leads to inaccuracy in measurement. Now we have proved that, at least at low opacity, the RI can be figured out without the interference from opacity by taking the power ratio between different diffraction orders. After we figure out the RI by power ratio, we can use the absolute power to figure out the opacity of the liquid.

**Conclusion**

We developed a multi-functional optofluidic sensor based on optical diffraction. In this work, we demonstrated two major improvements on this sensor by both simulation and experiment. One improvement is shifting and broadening the dynamic measurement range by making the grating using the polymer with higher refractive index. The other improvement is that the RI sensing and opacity sensing can be separated by taking the power ratio between different diffraction orders for RI sensing then using the absolute power for opacity sensing.

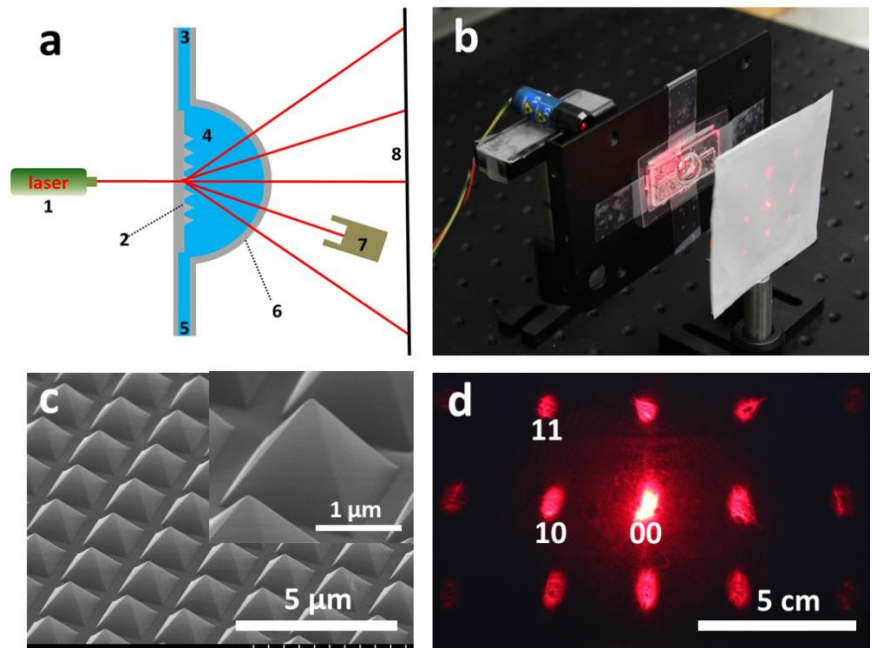

**Fig .1**. (a) Schematic of the diffraction sensing setup. ① Laser diode, 633 nm, 4mW, ② 2D pyramids diffraction grating, ③ fluid inlet, ④ liquid for testing, ⑤ fluid outlet, ⑥ hemispherical fluid chamber, ⑦ photo detector ⑧ Transmission diffraction screen. (b) Photograph of the diffraction sensing system. (c) SEM of the 2D pyramids diffraction grating. (d) Photograph of an exemplary transmission diffraction pattern in measurement.

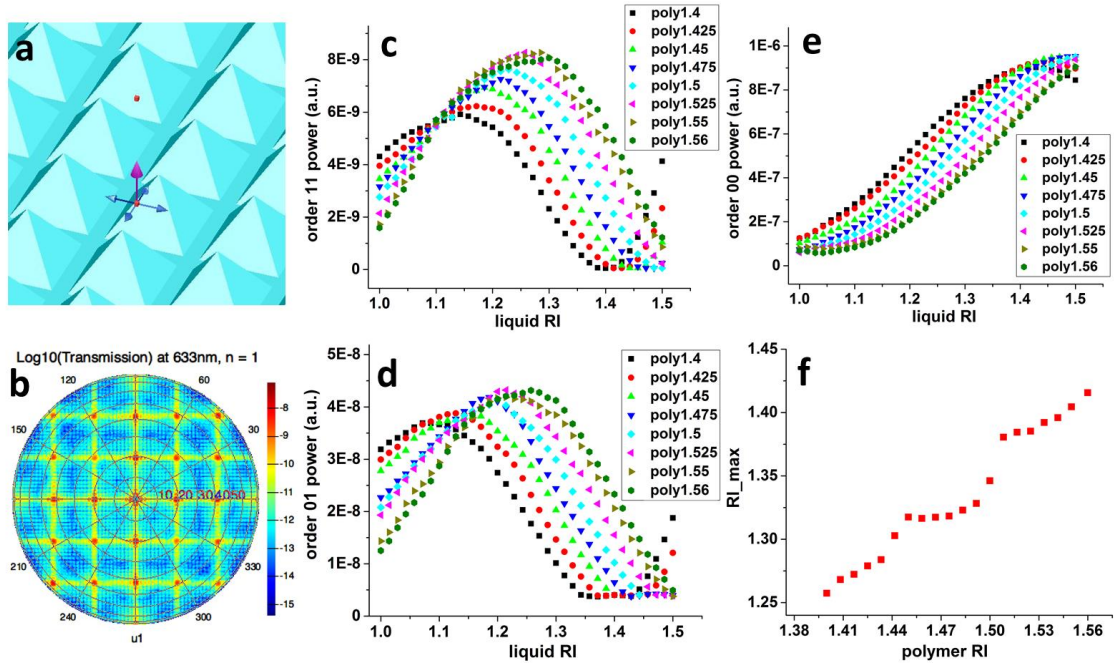

**Fig. 2**. (a) Pyramids array in FDTD simulation. The pink arrow indicates the direction of Poynting vector while the blue arrows indicate the directions of polarizations. (b) Simulated transmission diffraction pattern in the far field when RI of the environment is 1. (c-f) Simulation results for transmission diffraction power of order 11(c), order 10(d) and order 00(e) at different RI of liquid (x-axis) and grating (by different curve). (f) The most sensitive region for sensing of RI of liquid at different RI of grating.

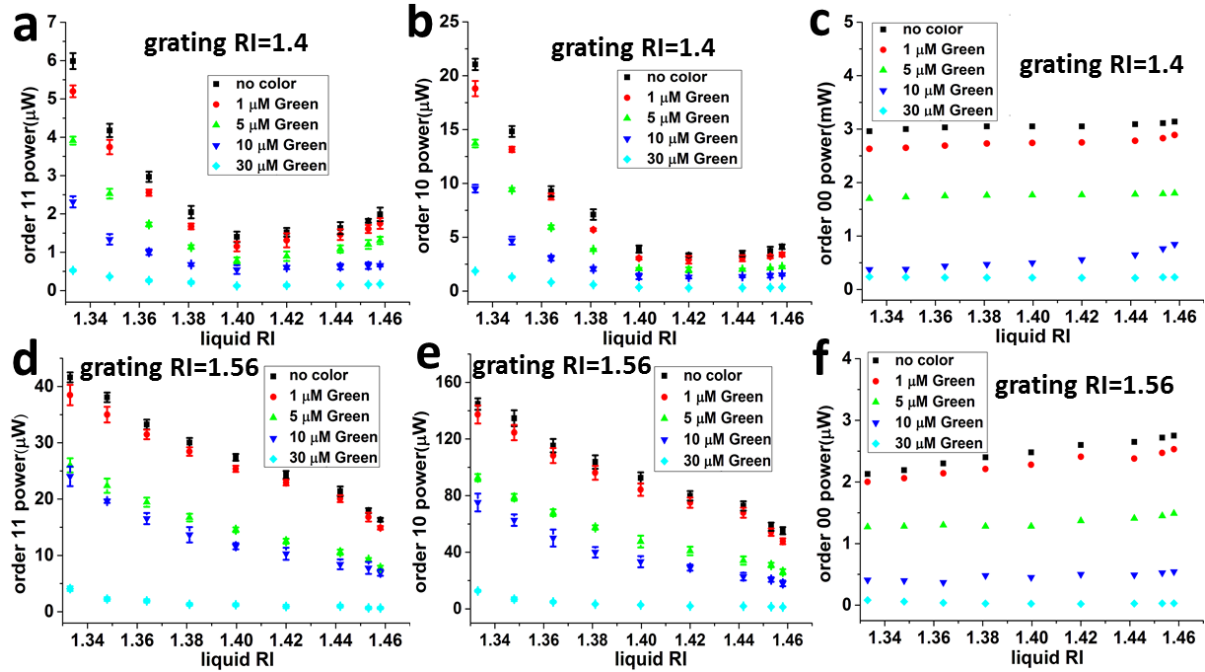

**Fig.3**. Measured results of diffraction power using liquid with different RI (concentration of sucrose) and absorption (concentration of brilliant green). Dots represent the mean values, errorbars represent the standard deviation of multiple measurements. (a-c) Results with grating RI = 1.4(PDMS), power of order 11(a), order 10(b) and order 00(c) changing with liquid RI(x-axis) and different absorptions. (d-f) Results with grating RI = 1.56 (NOA 61), power of order 11(a), order 10(b) and order 00(c) changing with liquid RI(x-axis) and different absorptions. Black dots stand for colorless solution, red dots stand for 1 μM brilliant green, green dots stand for 5 μM brilliant green, blue dots stand for 10 μM brilliant green and cyan dots stand for 30μM brilliant green.

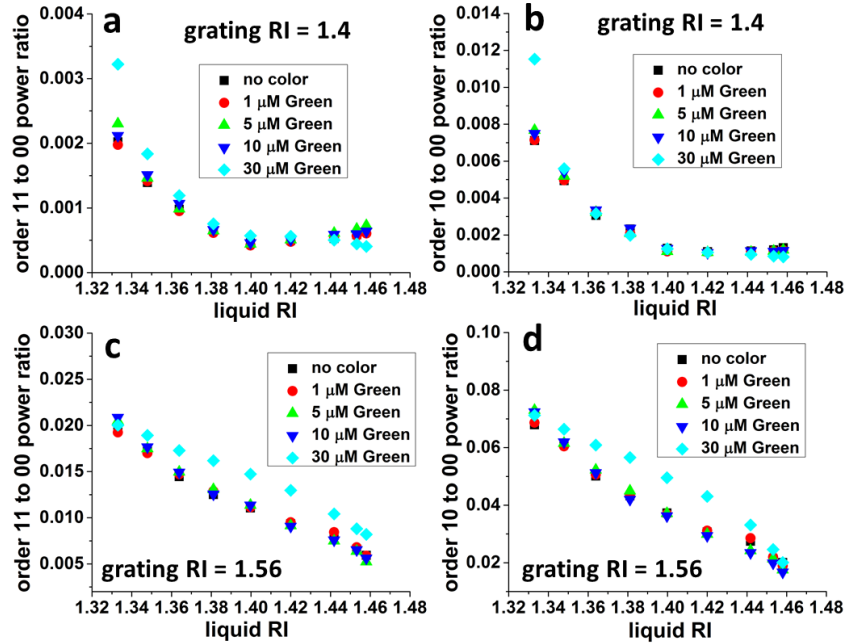

**Fig. 4**. Power ratio of diffraction orders using mean values of measurements in Fig. 3. Power ratio of order 11 to order 00 (a) and order 10 to order 00 (b) with grating RI = 1.4 (PDMS). Power ratio of order 11 to order 00 (c) and order 10 to order 00 (d) with grating RI = 1.56 (NOA 61).